\documentclass[10pt,twocolumn,letterpaper]{article}

\usepackage[pagenumbers]{iccv} 

\usepackage{makecell}
\usepackage{multirow}
\usepackage{multicol}

\definecolor{iccvblue}{rgb}{0.21,0.49,0.74}
\usepackage[pagebackref,breaklinks,colorlinks,allcolors=iccvblue]{hyperref}
\usepackage{algorithm}
\usepackage{algpseudocode}
\usepackage{amsmath}
\usepackage{amssymb}
\usepackage{geometry}
\usepackage{caption}
\usepackage{subcaption}
\usepackage{float}
\usepackage{xcolor}
\usepackage{multirow}
\usepackage{multicol}
\usepackage{booktabs}
\usepackage{tabularx}
\usepackage{color}
\usepackage{threeparttable} 

\def\paperID{9891} 
\def\confName{ICCV}
\def\confYear{2025}

\title{ Global Position Aware Group Choreography using Large Language Model}

\author{Haozhou Pang$^*$, Tianwei Ding$^*$, Lanshan He$^*$, and Qi Gan\\
Soul AI, China\\
\\
}

\begin{document}

\twocolumn[{%
\renewcommand\twocolumn[1][]{#1}%

\maketitle

\begin{center}
    \centering
    \captionsetup{type=figure}
    \includegraphics[width=1.0\linewidth]{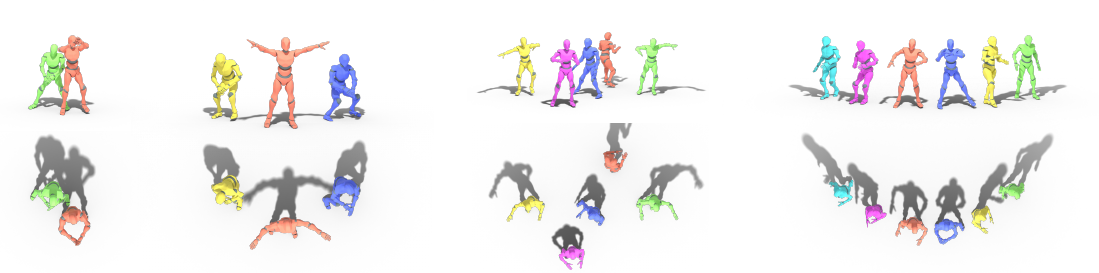}
    \caption{We present a framework to generate diverse and coherent group choreography using Large Language Model.}
\end{center}%
}]
\def\thefootnote{*}\footnotetext{These authors contributed equally to this work}\def\thefootnote{}\footnote{Preprint.}

\begin{abstract}




Dance serves as a profound and universal expression of human culture, conveying emotions and stories through movements synchronized with music. Although some current works have achieved satisfactory results in the task of single-person dance generation, the field of multi-person dance generation remains relatively novel. In this work, we present a group choreography framework that leverages recent advancements in Large Language Models (LLM) by modeling the group dance generation problem as a sequence-to-sequence translation task. Our framework consists of a tokenizer that transforms continuous features into discrete tokens, and an LLM that is fine-tuned to predict motion tokens given the audio tokens.  We show that by proper tokenization of input modalities and careful design of the LLM training strategies, our framework can generate realistic and diverse group dances while maintaining strong music correlation and dancer-wise consistency. Extensive experiments and evaluations demonstrate that our framework achieves state-of-the-art performance. 
\end{abstract}    
\section{Introduction}
\label{sec:intro}

Dance is a profound and universal aspect of human culture, serving as a medium for expressing emotions and narratives through synchronized movements with music. Despite its significance, the automatic generation of dance poses substantial challenges due to its intricate temporal and spatial dynamics. Similar to other generative tasks, music-to-dance synthesis has also been widely studied \cite{li2021learn,siyao2022bailando,tseng2022edge, li2024lodge, li2023finedance}. With the rapid development of deep learning algorithms and the availability of more publicly accessible datasets, this field has made significant progress in recent years. The rapid development of automated dance generation frameworks has significantly influenced numerous downstream applications, including dance education \cite{FdiliAlaoui03072014, choreomaster2021}, automated choreography \cite{li2021learn}, and virtual idols in the metaverse \cite{Valle_P_rez_2021}. These technologies empower animators and content creators by enabling them to take advantage of powerful AI capabilities to enhance efficiency and inspire creativity.


In this work, we present a framework for generating realistic group dance with high group correlation conditioned on music. As a more generalized task compared to solo dance synthesis, group choreography presents greater application potential but also faces unique complexities that extend beyond individual kinematics. Our method quantizes data from different modalities into tokens so that a pretrained LLM can be adapted to solve the motion generation task as a sequence-to-sequence translation problem. In summary, our contributions are the following.

\begin{itemize}
    \item We build our framework based on quantizers and LLMs to generate group dances according to input musics. Our method outperforms prior works on existing evaluation metrics and user studies.
    \item By integrating global position guidance for group choreography into the training framework, our approach demonstrates superior formation preservation and group consistency compared to prior methods, evidenced by substantial improvements in quantitative metrics and visual effects.
    \item Our framework can generate group dance of arbitrary length without being affected by accumulated errors, due to the special design of global position tokens in the training and inference phase of our framework.  

\end{itemize}

\noindent Our work is best enjoyed accompanied by the video demos.

\section{Related Work}
\label{sec:relatedwork
}

\subsection{Human Motion Synthesis and Music to Dance}
Synthesizing realistic 3D human motions is an essential task in various fields, including, but not limited to, games, films, robotics, and virtual reality applications. Human motion synthesis is an interesting topic that has been studied extensively. Early works use rule-based or graph-based methods \cite{10.1145/311535.311539, 10.1145/3596711.3596788, 10.1145/566654.566605} to synthesize motions, which mainly rely on a carefully handcrafted heuristic to map input modality to a set of motion nodes. For music to dance, extra care is needed to satisfy the music rhythmic constraints when designing such rules. Even though such methods are highly explainable and controllable, the bottleneck is also obvious. The generation diversity and naturalness is limited by the design of rules, and adding more motion units requires extra manual work, making such methods inappropriate for in-wild scenarios. In recent years, deep learning-based methods have drawn much attention, as they synthesize motions from implicit representation of training datasets, hence can be trained in an end-to-end manner. The development of leaning based music to dance generation frameworks is analogues to that of other generation tasks. We briefly introduce some representative works using different network architectures. Deterministic models including MLP \cite{Kucherenko_2020}, CNN \cite{10.1145/3596711.3596789}, RNN\cite{GrooveNet,huang2023dancerevolutionlongtermdance,yalta2019weaklysuperviseddeeprecurrent, 10.1145/3240508.3240526} and transformers \cite{li2020learninggeneratediversedance,li2021learn,li2023danceformermusicconditioned3d} have been studied. Such deterministic methods tend to produce mean poses due to the one-to-many nature of the dance generation task and tend to generate unrealistic dances due to the lack of proper restrictions to keep the generated pose within the specific domain of dance. To alleviate this problem, generative models have been implemented. For example, VAEs \cite{hong2022avatarclipzeroshottextdrivengeneration, han2024enchantdanceunveilingpotentialmusicdriven}, VQVAEs \cite{siyao2022bailando}, flow-based models \cite{Valle_P_rez_2021}, and diffusion-based models \cite{li2023finedance,tseng2022edge, li2024lodge, yang2023longdancedifflongtermdancegeneration}, have been studied in previous works. Common issues in motion generation tasks, such as the foot skating problem and the long-term freezing problem, have also been identified and studied in previous works \cite{yang2023longdancedifflongtermdancegeneration, li2024lodge}. We direct our readers to \cite{zhu2023humanmotiongenerationsurvey} for a comprehensive survey on motion generation tasks and their corresponding progresses.  

\subsection{Group Choreography}
Although some recent works have achieved satisfactory results in the task of single-person dance generation, the field of multi-person dance generation remains relatively novel. Multi-person dance generation is a more challenging task because, in addition to maintaining the naturalness and continuity of the generated dance, we must also consider the coordination among different dancers and the consistency of the overall dance. Certain tricks used in single-person dance generation, such as normalization of root motion trajectories, are not applicable to the multi-person dance generation scenario, and extra effort is needed to ensure dancer-wise trajectories consistency. The multi-person dance generation task has received relatively less attention in previous research, primarily due to the lack of high-quality multi-person dance datasets in the public domain. Le \textit{et al.} \cite{aiozGdance} proposed a multi-dancer dataset consisting of 16.7 hours of paired music and 3D motion from in-the-wild videos, covering 7 dance styles and 16 music genres.

\subsection{Human Motion Generation using LLMs}
With the rapid advancement of LLMs, numerous academic studies have leveraged these models to achieve breakthroughs in various domains. For example, AnyGPT \cite{zhan2024anygptunifiedmultimodalllm} and M3GPT \cite{luo2024m3gptadvancedmultimodalmultitask} have illustrated that a LLMs can integrate different modalities, such as text, audio, and images, to facilitate any-to-any multimodal interactions. Similarly, MotionGPT \cite{jiang2024motiongpt} has demonstrated that human motion can be treated as a specific language, allowing relevant tasks to be addressed using a LLMs. Our method adopts 
a similar modeling paradigm by using a pretrained LLMs to tackle the problem of multi-person dance generation conditioned on music. However, there are several key differences between our framework and previous work. First, we formulate the group dance generation task as a multi-turn dialogue process, enabling the model to perceive other dancers' movements when generating actions for new dancers. This design ensures coordinated movements between dancers, leading to significant improvements in group coordination metrics (FID) compared to baseline methods. Furthermore, we incorporate enriched global positional information during the training and inference process to address dancers' relative positioning in group choreography. This enhancement enables our model to better maintain formation structures and achieves a significantly reduced collision probability between dancers.

\section{Method}
\begin{figure*}
    \centering
    \includegraphics[width=1\linewidth]{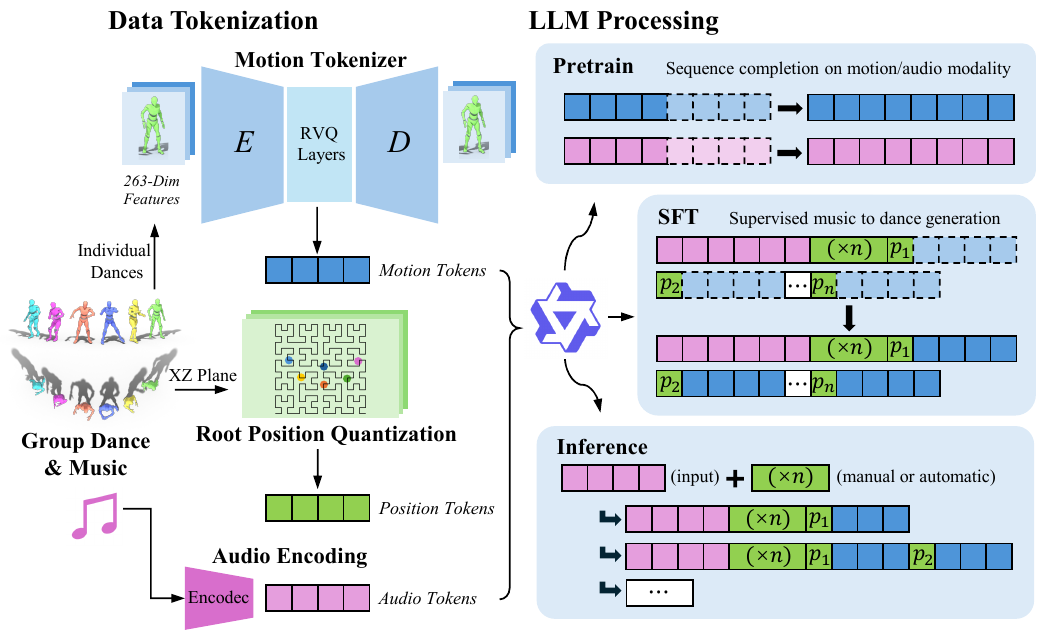}
    \caption{\textbf{Framework overview.} Our method consists of data tokenization and LLM processing. We transfer motions, global root positions, and audios into discrete tokens, respectively. After that, we carefully design the prompts and do LLM pretrain and tuning.}
    \label{fig:overview}
\end{figure*}
\noindent
\textbf{Problem Formulation.} Given an input music clip as a sequence of $\{a_1, a_2, \cdots, a_T\}$ where $t \in [1, T]$ is the index of the music segment, and the number of dancers $N$, our goal is to generate a set of motion sequences $\{m_1^1, \cdots, m_F^1; \cdots; m_1^{N}, \cdots, m_F^{N}\}$ where $m_f^i$ is the pose of the person $i$ at frame $f$.

\subsection{Motion and Music Tokenization}
While conventional approaches directly map continuous audio features to 3D pose sequences, we employ modality-specific tokenization to transform both audio and motion data into discrete symbolic representations via  codebook-based quantization. This paradigm shift fundamentally enhances information density by compressing continuous motion and audio  into compact discrete tokens, while simultaneously enabling compatibility with LLMs' sequence processing capabilities.

\noindent\textbf{Motion Tokenization.} We use the approach of Residual Vector Quantized Variational Autoencoder \cite{lee2022autoregressiveimagegenerationusing} (Residual VQVAE) to construct MotoinRVQ to tokenize the motion data. For a given motion sequence of a dancer $M = \{m_1, ..., m_F\}$. The MotionRVQ network can be represented as $Z=\Phi(M)$, where $\Phi(\cdot)$ is a function representing the MotionRVQ encoder. The quantization process is modeled as $Z^\ast = \sum_{l=1}^Lq_l(e_l)$ where $L$ is the number of quantization layers, $q_l(\cdot)$ is the quantization function at level $l$, and $e_l$ demotes the residual encoding at level $l$. The quantization at each level is defined as:

\begin{equation}
    q_l(e_l) = \mathop{\mathrm{arg\,min}}_{c \in C_l} \, \| e_l - c \|^2,
\end{equation}

where $C_l$ is the codebook at level $l$, and $c$ represents the embeddings in the codebook. The residual encodings are computed recursively as follows:
\begin{equation}
    \begin{aligned}
        e_1 &= \Phi(M), \\
        e_l &= e_{l-1} - q_{l-1}(e_{l-1}), \quad \text{for } l = 2,\dots,L.
    \end{aligned}
\end{equation}
We define the loss function to balance the reconstruction accuracy and codebook utilization. The total loss function $\mathcal{L}$ consists of three components:
\begin{equation}
    \mathcal{L} = \mathcal{L}_{rec} + \mathcal{L}_{commit} + \mathcal{L}_{codebook},
\end{equation}
where $\mathcal{L}_{rec} = \|M-\bar{M}\|_2^2$ is the reconstruction loss measuring the discrepancy between the original motion sequence and the reconstructed sequence;  $\mathcal{L}_{commit} = \beta \sum_{l=1}^{L}\|e_l - sg[q_l(e_l)]|_2^2$ is the commitment loss to ensure the encoder's outputs commit to the codebook entries;  $\mathcal{L}_{codebook} = \sum_{l=1}^L\|sg[e_l] - q_l(e_l)\|_2^2$ is the codebook loss to update the codebook entries to match the encoder outputs. By optimizing this loss function, the model learns to reconstruct motion sequences accurately while effectively utilizing the quantization codebooks. Training tricks including exponential moving average (EMA) and random re-initialization of inactivate codebook entries are used to ensure stable training process.

\noindent\textbf{Motion Representation.}
Following Guo et al.\cite{Guo_2022_CVPR}, we define a pose $m$ by a tuple of $(r^a, r^x, r^z, r^y, j^p, j^v, j^r, c^f)$, where $r^a$ is the root angular velocity along Y-axis; $(r^x, r^z \in \mathbb{R})$ are root linear velocities on XZ-plane; $r^y \in \mathbb{R}$ is root height; $(j^p, j^v \in \mathbb{R}^{J\times 3})$ are the local joints positions and velocities, $j^r \in \mathbb{R}^{J\times 6}$ are local joints rotations, where $J$ denotes the number of joints. $c^f \in \mathbb{R}^4$ is a binary vector that represents the foot-ground contacts.  It is observed that such pose representation contains redundant information since the joints' positions can be determined by forward kinetic calculation. However, we empirically find that such redundant representation is essential for a stable and high-quality tokenizer training. MotionRVQ network that trained solely on joint rotations results in  worse reconstruction quality and appeal to suffer from jitter artifacts. Noticing that only the velocity of root is considered in the pose representation, for each dancer, we additionally introduce $x \in \mathbb{R}^3$ to define the initial position of motion sequence.

\noindent\textbf{Audio Tokenization.} We use encodec \cite{defossez2022highfi}, a strong pretrained audio codec with quantized latent space, to perform  audio tokenization.

\subsection{Music-Driven Group Dance Generation Based on LLMs}

The audio and motion tokenizers convert low-dimensional, redundant raw data into discrete, expressive, and more compact representations, enabling further fine-tuning and alignment using LLMs.

\noindent \textbf{Phase 1: Cross-Modal Pretraining.} To enhance the LLM's understanding of motion and audio, we first train the model for next token prediction on motion tokens and audio tokens. Specifically, each motion label is converted into a word ``$\langle \text{motion\_id\_x} \rangle$", and each audio label is transformed into a word ``$\langle \text{music\_id\_x} \rangle$". By converting the sequence of motion labels and audio labels into text, we obtain a ``motion segment" and an ``audio segment," which are then encoded using the LLM's tokenizer to achieve modal pretraining. To improve the generalization and diversity of the LLM, we incorporate single-person dance data augmentation during this process.

\noindent \textbf{Phase 2: Supervised Fine-Tuning on Audio and Motion Modalities.} After obtaining the cross-modal pretrained model, we convert raw data  into ``segments" that LLMs can understand. Using the audio segments and motion segments, we construct text-based inputs and perform supervised fine-tuning by computing the loss on the motion segments. The supervised fine-tuning (SFT) objective is defined as:

\begin{equation}
\mathcal{L}_{\text{SFT}} = -\sum_{i=1}^{N}\sum_{t=1}^{T} \log P(m_{i, t} | m_{i, <t}, M_{<i}, A),
\end{equation}

where $m_{i,t}$ denotes the motion token of $i$-th dancer at time step $t$, $m_{i, <t}$ represents the motion tokens of $i$-th dancer before time $t$, $M_{<i}$ represents the motion token sequences of previous dancers, and $A$ denotes the audio token sequence.

\subsection{Global Position-Based Prompt Construction}

Multi-person dance generation requires coordination, primarily in terms of global positions and the synchronization of actions between characters. Global positions and character actions exhibit strong correlations. To better supervise the coordination of multi-character actions, we design a global position-guided mechanism.

\noindent \textbf{Global Position Quantization.} Given a character's spatial position, it is projected into the XZ plane to obtain coordinates $(x, z)$. Then we discretize them to position tokens as following:

\begin{equation}
\text{Pos\_id} = H(x, z),
\end{equation}

where $H(x, z)$ maps the 2D coordinates $(x, z)$ to a 1D discrete representation using the Hilbert curve.

\noindent \textbf{Training Phase Strategy.} Given an audio segment, the root positions of $N$ characters are mapped to discrete representations and converted into LLM-specific words ``$\langle \text{Pos\_id\_xx} \rangle$.'' All position words are concatenated after the audio segment, and the position word for each character is provided as a prompt before generating their actions. This enables the LLM to incorporate an understanding of global character positions.

\noindent \textbf{Inference Phase Strategy.} As shown in Algorithm (\ref{alg:inf}), long audio sequences are divided into segments during inference. We explain and demonstrate the inference effect of long audio in the supplementary materials. The initial position can be provided as a prompt by mapping specified coordinates to the Hilbert curve or by automatically selecting an initial position. After generating each segment, the end position of the motion is calculated and used as the initial position prompt for the next segment. This approach mitigates the accumulation of root position errors in long audio sequences and improves motion coordination.

\begin{algorithm}
\caption{Hierarchical Motion Generation with Positional Guidance}
\begin{algorithmic}[1]
\State \textbf{Input:} Audio sequence $\boldsymbol{A}$, Initial coord. $p_0$ and Hilbert map $H(\cdot)$
\State $N \gets \text{SegmentCount}(\boldsymbol{A})$ 

\For{$k\leftarrow 1$ to $N$}
    \State $\boldsymbol{M}_k \gets G_\theta(\boldsymbol{A}_k, H(p_{root}))$ 
    \State $\Delta p \gets \Phi(\boldsymbol{M}_k[-1])$ 
    \State $p_{root} \gets p_{root} + \Delta p$ 
\EndFor

\State $\boldsymbol{\hat{M}} \gets \text{MotionCodeMerge}(\{\boldsymbol{M}_k\}_{k=1}^N)$
\State $\boldsymbol{M} \gets 
\Phi(\boldsymbol{\hat{M}}$)

\State \textbf{Output:} Motion sequence $\boldsymbol{M}$ 
\end{algorithmic}
\label{alg:inf}
\end{algorithm}

\section{Experiments}

\subsection{Dataset}
We use the AIOZ-GDance dataset\cite{aiozGdance}, which contains a large amount of group dance data. It consists of 1,624 paired group dance motion and music clips in various dance styles and music genres. We keep consistent with the existing training-testing split as GDance\cite{aiozGdance}.\par

\subsection{Implementation Details}

\noindent \textbf{Motion RVQVAE.} The vector quantization module employs a residual quantization framework implemented through a hierarchical codebook architecture. The encoder-decoder structure integrates temporal 1D ResNet blocks and a 6-layer transformer backbone. Residual vector quantization incorporates 4 cascaded quantizers sharing a 512-entry codebook with 512-dimensional embeddings, where each quantizer progressively refines the latent residual from preceding stages. Codebook initialization leverages k-means clustering applied to the initial training batches, while exponential moving average updates ($\gamma$=0.95) stabilize codebook learning during training.

The training objective combines Smooth L1 reconstruction loss (0.8 weight) with vector quantization commitment loss (0.1 weight) and orthogonal regularization (0.1 weight) to prevent codebook collapse. 

\noindent\textbf{Finetuning Settings.} The model undergoes finetuning for 2 epochs using a global batch size of 32 distributed across 8 GPUs with automatic mixed precision (bfloat16). We employ the AdamW optimizer with base learning rate $2\times10^{-5}$ following a linear warmup over 500 steps, coupled with weight decay (0.01) and dropout (rate=0.1) regularization. Training operates on 2048-token sequences processed with gradient accumulation every 4 steps, monitored through automatic loss scaling and gradient clipping at norm 1.0.

The training of MotionRVQ Model cost about 15 hours with 1 L20 GPU.The finetuning of LLM (3B) cost about 4 hours with 8A100 GPUs.
\section{Evaluation}
\begin{table*}[h]
    \centering
    \begin{threeparttable} 
    \begin{tabularx}{\textwidth}{>{\raggedright\arraybackslash}X>{\centering\arraybackslash}X >{\centering\arraybackslash}X>{\centering\arraybackslash}X >{\centering\arraybackslash}X >{\centering\arraybackslash}X >{\centering\arraybackslash}X >{\centering\arraybackslash}X }
    \toprule[2pt]

  \multirow{2}{*}{\textbf{Methods}}  & \multicolumn{3}{c}{\textbf{Group}}  & \multicolumn{3}{c}{\textbf{Individual}}& \multicolumn{1}{c}{\textbf{User Study}}
  \\

  \cmidrule(lr){2-4} \cmidrule(lr){5-7} \cmidrule(lr){8-8}

   & \textbf{FID}$\downarrow$ & \textbf{Div}$\rightarrow$ &\textbf{TIF} $\downarrow$ &\textbf{FID}$\downarrow$  & \textbf{Div}$\rightarrow$ & \textbf{BA}$\uparrow$ & \textbf{Win Rate}
    \\

    \midrule

     \textbf{GT} & - & 10.63 & 0.092  &  - & 9.51 & 0.360 &  39.53\%  \\

     \textbf{LODGE\cite{li2024lodge}}    & 341.41 & 5.58 & 0.004\tnote{*} & 123.76 & 4.41 & 0.356  & 72.09\% \\

     \textbf{Bailando\cite{siyao2022bailando}} & 163.57 & 4.65& 0.228 &  112.12 & 4.60 & 0.341 & 86.05\%  \\

     \textbf{Ours}  & \textbf{42.79} & \textbf{6.62} & 0.102  &  \textbf{36.06} & \textbf{6.56} & 0.341 &  - \\

    \bottomrule[2pt]
    \end{tabularx}
    \end{threeparttable}\caption{Performance Comparison. This table presents quantitative metrics, including FID, Diversity, and TIF for various methods. The user study on the right side includes qualitative results obtained through anonymous sampling without replacement, comparing our method to others. Our approach demonstrates competitive quantitative performance alongside promising qualitative assessment, indicating its overall effectiveness. * LODGE exhibits little root movement in the results, so the TIF metric is significantly low.}
   \label{tab:metrics_comparison}
\end{table*}

\begin{figure}[hpbt]
    \centering
    \includegraphics[width=1\linewidth]{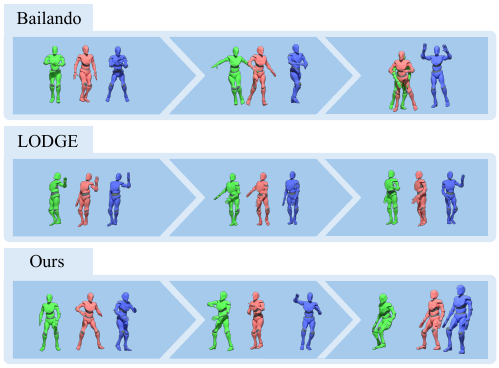}
    \caption{Visualization of different methods. Bailando tends to generate dance with cross-body intersection problem. Lodge generates dance with constrained root movements, resulting in less diverse group formation. Our method, empowered by global position guidance, enables more diverse formation patterns while significantly reducing character collision probabilities. For additional visual comparisons, please refer to the supplementary videos.}
    \label{fig:methods_compare}
\end{figure}

\subsection{Quantitative Metrics and Quality Comparison}
Following previous practices, we use the quantitative metrics below: Frechet Inception Distance (FID) in group features and in individual features, Generation Diversity, Beat Alignment Score(BA), and Trajectory Intersection Frequency(TIF). Specifically, FID measures the similarity between the generated and the ground-truth motion features; BA calculates the distances between audio beats and motion beats; TIF measures the frequency when a character collides with another. We use kinetic features proposed by AIST++ \cite{li2021learn} for individual FID and Diversity calculation. \par
Our evaluation  focuses on methodologically comparable approaches with fully reproducible implementations. We compare our metrics with LODGE and Bailando, two representative methods for generating music-conditioned single dancer motion, using diffusion-based and autoregression-based architectures, respectively. To generate a group's dance with these two methods, we repeat the inference process several times from random initial states. The results are listed in Table~\ref{tab:metrics_comparison}. It shows that our method not only achieves a better FID for individual features but also significantly excels at group features. \par

A user study is conducted for the quality comparison among the results of LODGE, Bailando and our method. There are 10 rendered video clips for each method, with the same start time and duration. We invite 43 users to rate these videos over three aspects: naturalness of each single character's motion, relevance between the music and the motions, and coordination of the group's motions. More details of user studies can be found in supplementary file. \par

\subsection{Ablation Study}

\textbf{Experiments on Base Models of Various Scales.} 
To investigate the impact of model scale on our task, we conducted experiments with Qwen2.5 \cite{qwen2.5} series models ranging from 0.5B to 7B parameters. We observed modest improvements in FID-related metrics when scaling from 0.5B to 3B parameters, but significant deterioration occurred when further increasing to 7B. These results suggest that larger models cannot achieve better performance given our current data scale. According to results from Hoffmann et al.\cite{10.5555/3600270.3602446}, the empirically optimal token-to-parameter ratio for LLM training should be approximately 20:1. This aligns with our findings, as our training data contains only approximately 24 million tokens, making smaller models more suitable. Interestingly, when comparing the original Qwen 1.5B model with the DeepSeek-R1-Distill \cite{deepseekai2025deepseekr1incentivizingreasoningcapability} counterpart under identical training configurations, the R1-distilled version showed better performance on FID metrics. We interpret this as evidence that foundation models demonstrating superior performance on text benchmarks possess inherent advantages for downstream tasks. See Table~\ref{tab:ablation_study} for details.

\begin{table}[hpbt]
    \centering
    \begin{tabular}{lcc}
    \toprule[1pt]
     \textbf{Model}  & \textbf{FID Group} & \textbf{FID Individual} \\
     \bottomrule
    \textbf{0.5B}    &  69.13 & \textbf{42.79}\\
     \textbf{1.5B}    &  76.16 & 49.90\\
     \textbf{1.5B R1}   & \textbf{65.47} &  48.13\\
     \textbf{3B}    &  77.56 & 53.16\\
     \textbf{7B}    &  94.44 & 71.65\\
     \bottomrule[1pt]
    \end{tabular}
    \caption{Experiments on Base Models of various scales. Our empirical analysis across diverse model scales demonstrates that, under the current dataset configuration, merely enlarging model parameters does not lead to proportional performance gains. However, enhanced model pretraining yields improvements in downstream evaluation metrics.}
    \label{tab:ablation_study}
\end{table}

\noindent 
\textbf{Effect of Pretrain.} Before SFT, pretraining on motion and audio  
helps the model to understand these modalities. We compare the pretrained models with those directly applied SFT. The results are listed in Table~\ref{tab:comparison_pretrain}.

\begin{table}[htbp]
    \centering
    \resizebox{\linewidth}{!}{
    \begin{tabular}{lcccc}
        \toprule[1pt]
        \multirow{3}{*}{\textbf{Model}} & \multicolumn{2}{c}{\textbf{FID Group}} & \multicolumn{2}{c}{\textbf{FID Individual}} \\
        \cmidrule(lr){2-3} \cmidrule(lr){4-5}
        & \makecell{w/o \\Pretrain}  & \makecell{w/ \\Pretrain} & \makecell{w/o \\Pretrain}   & \makecell{w/ \\Pretrain}  \\
        \midrule
        \textbf{0.5B} & 69.13 & \textbf{40.37} & 42.79 & \textbf{36.06} \\
        \textbf{1.5B R1} & 65.47 & \textbf{55.15} & 48.13 & \textbf{47.21} \\
        \textbf{3B} & 77.56 & \textbf{59.43} & 53.16 & \textbf{52.27} \\
        \bottomrule[1pt]
    \end{tabular}}
    \caption{Ablation study of the pretrain phase. All tested models achieve better FID metrics using the proposed two-phase training strtegy.}
    \label{tab:comparison_pretrain}
\end{table}

\par
\begin{figure}[hptb]
    \centering
    \includegraphics[width=0.8\linewidth]{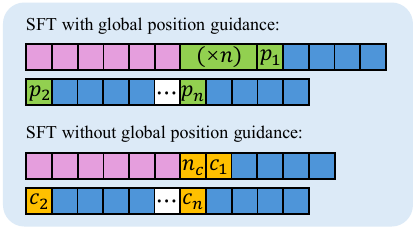}
    \caption{\textbf{SFT with/without Global Position Guidance.} In the prompt without global position guidance, a token \texttt{<}$n_c$\texttt{>} following the audio tokens indicates the total amount of characters, and there is an id token \texttt{<}$c_i$\texttt{>} for each character leading the motion tokens.}
    \label{fig:prompt_wo_pos}
\end{figure}

\noindent
\textbf{Effect of Global Position Guidance.} To prove the effectiveness of the global position guidances, we train LLM models without them, where the postion tokens are replaced with simple character tokens, as shown in Figure~\ref{fig:prompt_wo_pos}. 

By introducing the global position guidance, the generated group dances demonstrate enhanced spatial awareness, evidenced by improved formation preservation capability and reduced probability of inter-dancer collisions. As shown in Figure~\ref{fig:methods_compare}, the group's formation is better organized and there are fewer collisions, especially for long-time inference. We also quantitatively compare TIF between the models trained with and without global position guidance. The results in Table~\ref{tab:tif_on_pos_guidance} show that the probability of character collision has decreased after injecting the global position guidance among all models tested.


\begin{table}[hpbt]
    \centering
    \begin{tabular}{lcc}
    \toprule[1pt]
     \textbf{Model}  & \makecell{\textbf{TIF}\\w/o Position} & \makecell{\textbf{TIF}\\w/ Position} \\
     \midrule
     \textbf{0.5B}    &  0.182 & \textbf{0.102} \\
     \textbf{1.5B R1}    &  0.180 & \textbf{0.063} \\
     \textbf{3B }   &  0.158 & \textbf{0.104} \\
     \bottomrule[1pt]
    \end{tabular}
    \caption{Ablation study of the global position guidance.The experimental results indicate that after incorporating Position Guidance into base models of various sizes, the TIF metric decreases significantly.}
    \label{tab:tif_on_pos_guidance}
\end{table}
\section{Conclusion}
In this work, we study the problem of group dance generation conditioned on music. Compared to the well-studied single-dancer generation task, multi-dancer choreography demands stricter requirements for collective coordination and global dancer-wise consistency, making it a more challenging yet practical task with broader applications. Our method quantifies multi-modal input features into temporal-aligned tokens, reformulating group dance synthesis as a multi-turn dialogue framework that is subsequently fine-tuned on group dance data using LLMs. The proposed two-phase training strategy combined with global positional guidance in the training and inference process, further improves the overall generation quality and diversity of our framework, evidenced by a boost in the FID and diversity metrics. Evaluation on existing metrics along with user studies show that our framework surpasses previous methods in single-dancer metrics while achieving significant improvements in group metrics. 
{
    \small
    \bibliographystyle{ieeenat_fullname}
    \bibliography{main}
}

\end{document}